\documentstyle[epsfig,sprocl]{article}

\def\Journal#1#2#3#4{{#1} {\bf #2}, #3 (#4)}

\def\NPA{{\em Nucl. Phys.} A}
\def\PLB{{\em Phys. Lett.}  B}
\def\PRL{\em Phys. Rev. Lett.}

\def\PRC{{\em Phys. Rev.} C}

\def\PR{\em Phys. Rep.}

\def\be{\begin{equation}}
\def\ee{\end{equation}}
\def\bea{\begin{eqnarray}}
\def\eea{\end{eqnarray}}
\begin{document}

\title{DYNAMICS OF KAONS IN NUCLEAR MATTER
~\footnote{This talk is dedicated to
Mannque Rho on the occasion of his 60th birthday}}

\author{M.F.M. LUTZ}

\address{GSI, 64220 Darmstadt, Germany
\\E-mail: m.lutz@gsi.de}


\maketitle\abstracts{
We consider $K^-$ nucleon elastic and inelastic scattering in
isospin symmetric nuclear matter. It is found that the proper
description of the $\Lambda(1405)$ resonance structure in nuclear
matter requires a self-consistent approach. Then the $\Lambda
(1405) $ resonance mass remains basically unchanged, however, the
resonance acquires an increased decay width as nuclear matter is
compressed. The elastic and inelastic scattering amplitudes show
important medium modifications close to the kaon nucleon threshold.
We also construct the $K^-$-nuclear optical potential appropriate
for $K^-$ atoms. Our microscopic approach predicts a strong
attractive non-local interaction strength.}

\section{Introduction}

In this talk we report on recent work \cite{letter} on kaon
propagation and kaon nucleon interaction in isospin symmetric
nuclear matter. There has been much effort to evaluate the in
medium $K$-mass in realistic models. Gerry Brown and Mannque Rho
\cite{Brown} suggested an effective  mean field approach. Here we
focus on a microscopic description of the kaon propagator in dense
matter as derived from kaon nucleon scattering in free space
\cite{Yabu,Koch,Waas1,Waas2,Pethick,Min}.

Let us review the change of the $K^+$-mass. As was emphasized in
\cite{Lutz} it is given by the low density theorem in terms of the
empirical $K^+$-nucleon scattering lengths $a_{K^+N}^{(0)}\simeq
0.02 $ fm and $a_{K^+N}^{(1)}\simeq -0.32 $ fm \cite{Martin}
\begin{eqnarray}
\Delta \,m_{K^+}^2 = -\pi \left( 1+\frac{m_K}{m_N} \right)
\left( a^{(I=0)}_{K^+N}+3\,a^{(I=1)}_{K^+N} \right)
\,\rho +{\cal O}\left( k_F^4 \right)
\label{lowdensity}
\end{eqnarray}
where $\rho = 2\,k_F^3/(3\,\pi^2 )$. Model calculations
\cite{Min,Waas2} typically find only small corrections to
(\ref{lowdensity}). In fact the next to leading term of order
$k_F^4 $ can be evaluated model independently in terms of the $K^+
N$-scattering lengths
\begin{eqnarray}
\Delta \,m_{K^+}^2 =(1)+
\alpha \left( \left(a_{K^+N}^{(I=0)}\right)^2 +3\,\left(a_{K^+N}^{(I=1)}\right)^2 \right)
 k_F^4 +{\cal O}\left( k_F^5 \right)
\label{correction}
\end{eqnarray}
where
\begin{eqnarray}
\alpha &&=\frac{1-x^2+x^2\,\log
\left(x^2\right)}{\pi^2\,(1-x)^2}
\simeq 0.166
\label{}
\end{eqnarray}
and $x =m_K/m_N $. Note that one must not expand the correction
term (\ref{correction}) in powers of $m_K/m_N$. At $x=0 $ with
$\alpha \simeq 0.10 $ the real part of the correction term is
underestimated by more than $60 \%$.

In Fig. 1 the leading and subleading change of the $K^+$-mass as a
function of the Fermi momentum, $k_F$, is shown.
\begin{figure}[h]
\epsfysize=5cm
\begin{center}
\mbox{\epsfbox{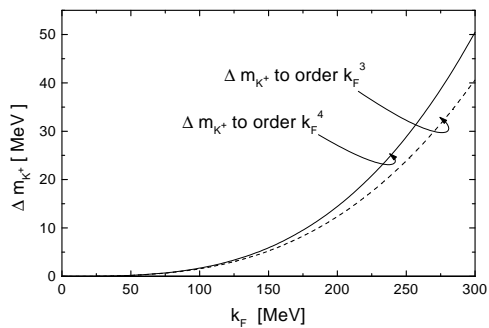}}
\end{center}
\caption{Change of $K^+$-mass in isospin symmetric nuclear matter.}
\label{fig1}
\end{figure}
The correction term is indeed small so that the density expansion
is useful in the $K^+$-channel. At nuclear saturation density with
$k_F \simeq 265 $ MeV it increases the repulsive $K^+$-mass shift
from $28$ MeV to $ 35 $ MeV by about $20 \%$. We conclude that any 
microscopic model consistent with low energy $K^+$-nucleon 
scattering data is bound to give similar results for the 
$K^+$-propagation in nuclear matter at densities $\rho 
\simeq \rho_0 $ sufficiently small to maintain the density expansion rapidly
convergent. Consequently the role of chiral symmetry is restricted
to the qualitative prediction of the $K^+N$-scattering lengths by
the Weinberg-Tomozawa term.

We turn to the $K^-$-mass. At nuclear saturation the density
expansion for the $K^{-}$-mode is poorly convergent if at all
\cite{letter}. The leading terms of the density expansion appear to
contradict kaonic atom data \cite{Gal,report} which suggest sizable
attraction at small density. Furthermore the empirical $K^-N$
scattering lengths $a_{K^-N}^{(0)}\simeq (-1.70+i\,0.68 )$ fm and
$a_{K^-N}^{(1)}\simeq (0.37 +i\,0.60)$ fm \cite{Martin,Iwasaki} are
in striking disagreement with the Weinberg-Tomozawa term, the
leading order chiral prediction.

The solution to this puzzle lies in the presence of the $\Lambda
(1405) $ resonance in the $K^-$-proton channel
\cite{Koch,Waas1,Kaiser}. The $\Lambda(1405)$-resonance can be
described together with elastic and inelastic $K^-$-proton
scattering data  in terms of a coupled channel Lippman-Schwinger
equation with the potential matrix evaluated perturbatively from
the chiral Lagrangian \cite{Kaiser}. However, the role of chiral
symmetry in this approach is unclear. The systematic treatment of
important chiral correction terms like range terms or loop effects
remains unsettled. Note that a satisfactory description of kaon
nucleon scattering data can also be achieved by the coupled
$K$-matrix approach of Martin \cite{Martin}.

As was pointed out first by Koch \cite{Koch}, the $\Lambda (1405 )$
resonance may experience a repulsive mass shift due to Pauli
blocking which strongly affects the in-medium $K^-$-nucleon
scattering amplitude \cite{Koch,Waas1}. This offers a simple
mechanism for the transition from repulsion, implied by the
scattering lengths, at low densities, $\rho < 0.1 \,\rho_0 $, to
attraction at somewhat larger densities \cite{Waas1}. Schematically
this effect can be reproduced in terms of an elementary $\Lambda
(1405 )$ field dressed by  a kaon nucleon loop. The repulsive
$\Lambda $ mass shift due to the Pauli blocking  of the nucleon is
given by:
\begin{eqnarray}
\Delta\,m_\Lambda =\frac{g_{\Lambda NK}^2}{\pi^2}\,\frac{m_N}{m_\Lambda} \,
\left(1-\frac{\mu_\Lambda}{k_F}\,\arctan \left( \frac{k_F}{\mu_\Lambda} \right)
\right)k_F
\label{lambda}
\end{eqnarray}
with the 'small' scale
\begin{eqnarray}
\mu_\Lambda^2 &=&\frac{m_N}{m_\Lambda }\left(m_K^2-\Big(m_\Lambda-m_N\Big)^2 \right)
\simeq \left( 144\, MeV \right)^2
\label{scale}
\end{eqnarray}
and the $\Lambda(1405)  $ kaon nucleon coupling constant
$g_{\Lambda NK}$.

From (\ref{lambda}) we conclude that it is important to extend
previous work \cite{Koch,Waas2} and treat the $K^-$-state and the
$\Lambda(1405)$ states self consistently \cite{letter}. We expect
self consistency  to be important for in-medium $K^-N$-scattering
simply because the characteristic scale $\mu_\Lambda $ in
(\ref{scale}) depends sensitively on small variations of $
m_\Lambda $ and $m_K$. The $\Lambda(1405) $-resonance mass in
matter is now the result of two competing effects: the Pauli
blocking increases the mass whereas the decrease of the $K^-$ mass
tends to lower the mass.

\section{ $K^-$ nucleon scattering}

We describe $K^-$ nucleon scattering by means of an effective
Lagrangian density:
\begin{eqnarray}
{\cal L} &=&{\textstyle{1\over 2}}\,g_{11}^{(I=0)}\,
\left( N^\dagger  \, K\right)
\left( K^\dagger \, N \right)
+{\textstyle{1\over \sqrt{6}}}\,g_{12}^{(I=0)}\,
\left(  N^\dagger \, \,K \right)
\left( \vec \pi^\dagger \cdot\vec \Sigma  \right)
\nonumber\\
&+&{\textstyle{1\over \sqrt{6}}}\,g_{21}^{(I=0)}\,
\left(  \vec \Sigma^\dagger \cdot \vec \pi \right)
\left(K^\dagger  \, N \right)
+{\textstyle {1\over 3}}\,g_{22}^{(I=0)}\,
\left(  \vec \Sigma^\dagger \cdot \vec \pi \right)
\left(  \vec \pi^\dagger \cdot \vec  \Sigma  \right)
\nonumber\\
&+&{\textstyle{1\over 2}}\,g_{11}^{(I=1)}\,
\left( N^\dagger  \,\vec \tau \, K\right)
\left( K^\dagger \,\vec \tau\, N \right)
-{\textstyle{1\over 2}}\,g_{22}^{(I=1)}\,
\left(  \vec \Sigma^\dagger \times \vec \pi \right)
\left(  \vec \pi^\dagger \times \vec \Sigma  \right)
\nonumber\\
&+& g_{33}^{(I=1)}\,
\left( \Lambda^\dagger \vec \pi \right)
\left( \vec \pi^\dagger \, \Lambda \right)
-{\textstyle{i\over 2}}\,g_{12}^{(I=1)}\,
\Big[ \left( N^\dagger \,\vec \tau \, K \right)
\left( \vec \pi^\dagger \times \vec \Sigma \right)-h.c. \Big]
\nonumber\\
&+&{\textstyle{1\over \sqrt{2}}}\,g_{13}^{(I=1)}\,
\Big[\left( N^\dagger  \,\vec \tau \, K\right)
\left( \vec \pi^\dagger \, \Lambda \right)+h.c. \Big]
\nonumber\\
&-&{\textstyle{i\over \sqrt{2}}}\,g_{23}^{(I=1)}\,
\Big[\left( \vec \Sigma^\dagger  \times \vec \pi   \right)
\left( \vec \pi^\dagger \, \Lambda \right) -h.c. \Big]
\label{lagrangianzero}
\end{eqnarray}
with the isospin doublet fields $K=(K_{-}^\dagger,\bar K_0^\dagger)
$ and $N=(p,n)$. Here we include  the pion, the $\Sigma( 1195 )$
and the $\Lambda(1115) $ as relevant degrees of freedom since they
couple strongly to the $K^{-}$-nucleon system. The baryon fields as
well as the kaon and pion fields are constructed with relativistic
kinematics but without anti-particle components. Technically one
represents the interaction strength generated by loops with
propagating anti kaons or anti nucleons by local 4-point
interaction terms. For the nucleon and kaon this is certainly a
'clean' procedure for energies not far from the kaon nucleon
threshold since the kaon and nucleon mass are sufficiently heavy.
Integrating out the anti particles of the pion (two-particle
irreducible pion loops) is more subtle since one might expect such
a scheme to be restricted to energies close to the $\pi \Sigma $-
threshold. Here we assume that the 'small-scale' non-localities
which originate from integrating out the anti pions  are not
important at energies close to the kaon nucleon threshold. Since
our scheme describes low energy elastic and inelastic $K^-$-nucleon
scattering data this appears to be justified a posteriori.

The free nucleon and kaon propagator take the form:
\begin{eqnarray}
S_N(\omega, \vec q \,) &=&\frac{m_N}{E_N(q)}\,
\frac{1}{\omega -E_N(q)+i\,\epsilon }
\nonumber\\
S_K(\omega, \vec q \,) &=&\frac{1}{2\,E_K(q )}\,
\frac{1}{\omega -E_K(q)+i\,\epsilon } \; ,
\label{}
\end{eqnarray}
respectively, where $E_a(q)=\sqrt{ m_a^2+q^2 } $. The isospin zero
coupled channel scattering amplitude
\begin{eqnarray}
T=
\left(
\begin{array}{cc}
T_{KN\rightarrow KN} & T_{KN\rightarrow \pi \Sigma } \\ T_{\pi
\Sigma \rightarrow KN } & T_{\pi \Sigma \rightarrow \pi \Sigma }
\end{array}
\right)
\label{}
\end{eqnarray}
is given by the set of ladder diagrams conveniently resummed in
terms of the Bethe-Salpeter integral equation. Since the
interaction terms in (\ref{lagrangianzero}) are local the
Bethe-Salpeter equation reduces to the simple matrix equation
\begin{eqnarray}
T(s)=g(s)+g(s)\,J(s)\,T(s) =\left( g^{-1}(s)- J(s) \right)^{-1} \;
.
\label{ts}
\end{eqnarray}
with the loop matrix $J=$ diag $\,( J_{KN}, J_{\pi
\Sigma} ) $ and
\begin{eqnarray}
J_{KN} (\omega, \vec q\,) &=&-\int_0^{\lambda } \frac{d^3
l}{(2\pi)^3 }\,
\frac{m_N}{E_N(l)}\,S_K(\omega
-E_N(l),\vec q-\vec l\,)
\nonumber\\
 J_{\pi \Sigma } (\omega ,\vec q\,) &=&-\int_0^{\lambda }
\frac{d^3 l}{(2\pi)^3 }\,
\frac{m_\Sigma}{E_\Sigma(l)}\,S_{\pi }(\omega
-E_\Sigma(l),\vec q-\vec l\,) \, .
\label{}
\end{eqnarray}
Small range terms are included in our scheme by the replacements $
g_{11} \rightarrow g_{11}+h_{11} \left( s- (m_N+m_K )^2
\right)$, $ g_{12}\rightarrow g_{12}+h_{12} \left( s- (m_N+m_K )^2 \right)
$ and $ g_{22}\rightarrow g_{22}+h_{22} \left( s- (m_\Sigma+m_\pi
)^2
\right) $ induced by appropriate additional terms in
(\ref{lagrangianzero}). The loop functions $J_{KN}$ and $J_{\pi
\Sigma }$ are regularized by the cutoff $\lambda =0.7$ GeV.

We construct the $I=1$ coupled channel scattering amplitude in full
analogy to the isospin zero case with $J=$ diag $(J_{KN},J_{\pi
\Sigma },J_{\pi \Lambda } )$ (see eq. (\ref{ts}) ). The range terms
are included by the replacements $ g_{11}
\rightarrow g_{11}+h_{11} \left( s- (m_N+m_K )^2
\right)$, $ g_{12}\rightarrow g_{12}+h_{12} \left( s- (m_N+m_K )^2 \right)
$, $ g_{13}\rightarrow g_{13}+h_{13} \left( s- (m_N+m_K )^2 \right)
$, $ g_{22}\rightarrow g_{22}+h_{22} \left( s- (m_\Sigma+m_\pi )^2
\right) $ and
$ g_{33}\rightarrow g_{33}+h_{33} \left( s- (m_\Lambda+m_\pi )^2
\right) $.

The Lagrangian density (\ref{lagrangianzero}) follows from a chiral
Lagrangian with relativistic baryon and meson fields upon
integrating out the anti-particle field components. Therefore the
coupling matrix $g$ is constrained to some extent by chiral
symmetry \cite{Kaiser}. To leading order one may  derive the
isospin zero coupling strengths by matching tree level threshold
amplitudes. The chiral matching of correction terms and the range
parameters $h_{ij} $ is less obvious. In fact a consistent chiral
matching requires the $K^-$-nucleon potential to be evaluated
minimally at chiral order $Q^3$. Only at this order the necessary
counter terms for the loop functions are introduced. Recall that,
as was emphasized by Kolck \cite{Bira}, only the cutoff dependent
real part of the loop functions trigger the desired sign change of
the $K^-$ proton scattering length.

In this work the coupling strengths $g_{ij}$  are directly adjusted
to reproduce empirical scattering data described in terms of the
coupled channel scattering amplitude (\ref{ts}). Here we might
loose some constraint from chiral symmetry. The effective range
parameters $h_{ij} $, on the other hand, we construct according to
the $SU(3)$-symmetry of the chiral Lagrangian since they are not
affected by the chiral $Q^3$ counter terms. This implies that the
range terms in the isospin one channel can be expressed in terms of
the isospin zero range parameters and one free parameter $h_F$
\begin{eqnarray}
h_{11}^{(I=1)}&=&{\textstyle{1\over2}}\,\left(h_{11}^{(I=0)}-\sqrt{6}\,h_{12}^{(I=0)}+h_{22}^{(I=0)}-6\,h_F\right)
\nonumber\\
h_{12}^{(I=1)}&=&{\textstyle{1\over6}}\,\left(3\,h_{11}^{(I=0)}+\sqrt{6}\,h_{12}^{(I=0)}-3\,h_{22}^{(I=0)}-6\,h_F\right)
\nonumber\\
h_{13}^{(I=1)}&=&{\textstyle{1\over12\sqrt{6}}}\,\left(6\,h_{11}^{(I=0)}+2\,\sqrt{6}\,h_{12}^{(I=0)}-6\,h_{22}^{(I=0)}+36\,h_F\right)
\nonumber\\
h_{22}^{(I=1)}&=&{\textstyle{1\over12}}\,\left(15\,h_{11}^{(I=0)}-5\,\sqrt{6}\,h_{12}^{(I=0)}-3\,h_{22}^{(I=0)}-30\,h_F\right)
\nonumber\\
h_{23}^{(I=1)}&=&0
\nonumber\\
h_{33}^{(I=1)}&=&{\textstyle{1\over9}}\,\left(3\,h_{11}^{(I=0)}-5\,\sqrt{6}\,h_{12}^{(I=0)}+6\,h_{22}^{(I=0)}-18\,h_F\right)
\label{su3ranges}
\end{eqnarray}
The imposed $SU(3)$-symmetry for the range parameters
(\ref{su3ranges}) is an important ingredient for our subthreshold
extrapolation of the $K^-N$-scattering amplitude in the $I=1$
channel. Note that the chiral approach of reference \cite{Kaiser}
starts with a SU(3) symmetric interaction. However, the use of
different cutoff values in the various loop integrals obscures the
SU(3)-symmetry constraint.

In the isospin zero channel the  set of parameters $
g_{11}\,\lambda
=46.86 $, $ g_{12}\, \lambda = 11.67 $, $ g_{22}\, \lambda = 16.08 $,
$ h_{11}\,\lambda^3 = 0.79 $, $ h_{12}\, \lambda^3 = 8.57$ and
$h_{22}\,\lambda^3 =4.94 $ follows from a least square fit to the
amplitudes of \cite{Kaiser}. The isospin one amplitudes $K^{-}N \rightarrow K^{-}N, \pi
\Sigma ,\pi \Lambda $ of \cite{Kaiser} are described well with $g_{23}
=g_{33}=0$  as suggested by the leading order Weinberg Tomozawa term
and $g_{11}\,\lambda =12.75 $, $g_{12}\,\lambda =13.56 $,
$g_{13}\,\lambda =15.07 $ , $g_{22}\,\lambda
= 16.04 $ and $h_F\,\lambda^3 =-1.92 $.

We obtain a good description of all coupled channel amplitudes with
the scattering lengths $a^{(I=0)}_{K^-N}\simeq (-1.76 + i\,0.60) $
fm and $a_{K^-N}^{(I=1)} \simeq (0.35 + i\, 0.69 ) $ fm close to
the values obtained by Martin \cite{Martin}.

%\newpage

\section{$K^-$ in nuclear matter}

The kaon self energy $\Pi_K(\omega ,\vec q\, ) $ is evaluated in
the nucleon gas approximation
\begin{eqnarray}
\Pi_K \left( \omega , \vec q\, \right)
&= &- 4\,\int_0^{k_F}\frac{d^3l}{(2\,\pi)^3}\,\frac{m_N}{E_N(l)}\,
 \bar T_{KN}(\omega
+E_N(l),
\vec q+\vec l\,)
\label{kaonself}
\end{eqnarray}
in terms of the in medium kaon nucleon scattering amplitude
$4\,\bar T_{KN}=\bar T_{KN}^{(I=0)}+3\,\bar T_{KN}^{(I=1)}$. The
scattering amplitude $\bar T_{KN}$ is given by (\ref{ts}) with the
vacuum kaon nucleon loop $J_{KN} $ replaced by the in matter loop
$\bar J_{KN} $ with
\begin{eqnarray}
\bar J_{KN} \left( \omega , \vec q\, \right)
&=& -\int_{k_F}^{\lambda }\frac{d^3l}{(2\,\pi)^3}\,
\frac{m_N}{E_N(l)}\,
\bar S_K(\omega-E_N(l),\vec q-\vec l\,)
\label{}
\end{eqnarray}
and the Fermi momentum $k_F$. The in medium kaon nucleon loop,
$\bar J_{KN}(\omega , \vec q\, ) $ is regularized by our cutoff
parameter $\lambda $ such as to reproduce the vacuum loop function
$J_{KN} (\omega , \vec q, ) $ in the zero density limit.
Selfconsistency is met once $\bar J_{KN} $ is evaluated in terms of
the kaon propagator:
\begin{eqnarray}
\bar S_K(\omega , \vec q \, )
&=& \frac{1}{2\,E_K(q) }\,\frac{1}{\omega -E_K(q)-\Pi_K(\omega ,
\vec q \,) /(2\,E_K(q)\,)+i\,\epsilon }
\label{}
\end{eqnarray}
with the kaon self energy $\Pi_K (\omega , \vec q \, ) $ of
(\ref{kaonself}).

\begin{figure}[h]
\epsfysize=8cm
\begin{center}
\mbox{\epsfbox{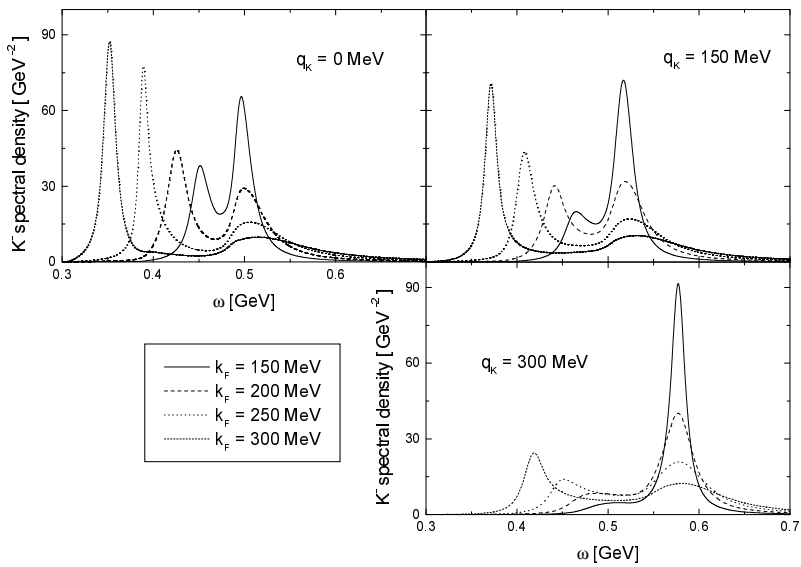}}
\end{center}
\caption{$K^-$ spectral density for kaon momenta $q_K=0$,
$q_K=150 $ MeV and $q_K=300$ MeV in isospin symmetric nuclear
matter. The spectral density is evaluated self consistently with
respect to the propagation of the $\Lambda (1405) $ resonance.}
\label{fig2}
\end{figure}

In Fig. 2 we present our final self-consistent result for the kaon
spectral density as a function of the kaon energy $\omega $ for
various Fermi momenta $k_F$ and kaon momenta $\vec q $. Typically
the spectral density exhibits a two peak structure representing the
$K^-$ and the $\Lambda (1405) $-nucleon hole states. As the Fermi
momentum $k_F$ increases the energetically lower state experiences
a strong attractive shift whereas the more massive state becomes
broader. On the other hand as the kaon momentum increases both
states basically gain kinetic energy with the energetically higher
peak attaining more strength.

It is instructive to compare various approximations for the kaon
self energy:  the kaon self energy as evaluated i) with the Pauli
blocked kaon nucleon scattering amplitude ( the approximation
scheme applied in \cite{Koch,Waas1,Waas2} ), ii) with the self
consistent amplitude and iii) with the  free space amplitude. All
three schemes predict a two mode structure of the kaon spectral
density, however, with quantitative differences. Both, the Pauli
blocked and the self-consistent amplitude shift strength from the
upper branch to the lower branch as compared with the spectral
density derived from the free space amplitude. While the Pauli
blocked amplitude causes a small repulsive shift of the $\Lambda
(1405)$--nucleon-hole state the self-consistent amplitude predicts
an attractive shift. At larger densities self consistency affects
the $K^-$-rest mass moderately. For example at $k_F=300 $ MeV we
find $\Delta \, m_{K^-}
\simeq -140 $ MeV and $\Gamma_{K^-} \simeq 35 $ MeV as compared with
$\Delta \, m_{K^-}
\simeq -141 $ MeV and $\Gamma_{K^-} \simeq 31 $ MeV from the free space
amplitude and $\Delta \, m_{K^-} \simeq -123$ MeV and $\Gamma_{K^-}
\simeq 29$ MeV from the Pauli-blocked amplitude. Note here that
the quasi particle width $\Gamma_{K^- }= -\Im
\,\Pi (m_{K^-}+\Delta\, m_{K^-}, \vec q =0 )/(m_{K^-}+\Delta \, m_{K^-} )$,
given above, differs from the physical $K^-$-width by about a
factor of two due to the strong energy dependence of the kaon self
energy (see Fig. 2). We find that self consistency is most
important once the $K^-$-mode starts moving relative to the nuclear
medium. Here we expect p-wave $K^-N$-interactions
\cite{Kolomeitsev}, which are not included in this work, to modify
our results to some extent.

\begin{figure}[h]
\epsfysize=8cm
\begin{center}
\mbox{\epsfbox{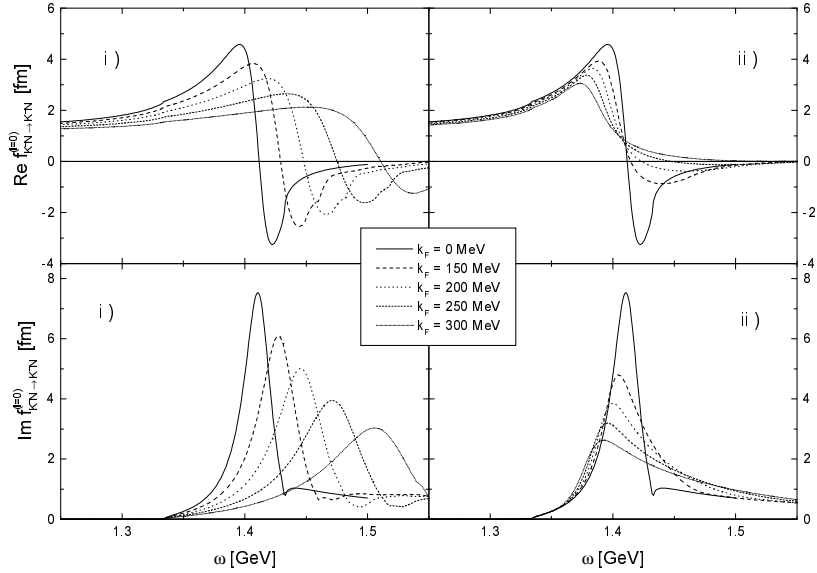}}
\end{center}
\caption{Isospin zero elastic $K^-$-nucleon scattering amplitude
in nuclear matter evaluated i) with Pauli blocked kaon nucleon loop
and ii) with self-consistent loop. Here $q_N+q_K=0 $.}
\label{fig3}
\end{figure}

We turn to in medium $K^-$ nucleon scattering. Fig. 3 shows the
isospin zero elastic scattering amplitude
\begin{eqnarray}
f^{(I=0)}_{K^-N\rightarrow K^-N }(\omega, \vec q\,)=
\frac{m_N}{4\,\pi
\,\omega }
\,T^{(I=0)}_{K^-N\rightarrow K^-N } (\omega,\vec q\, )
\label{}
\end{eqnarray}
evaluated with i) the Pauli blocked kaon nucleon loop and ii) the
self-consistent kaon nucleon loop at $|\vec q_K+\vec q_N| =0 $. The
two amplitudes differ dramatically as the Fermi momentum is
increased. In the Pauli-blocked amplitude i) the $\Lambda (1405)$
resonance peak is shifted to higher energies whereas the self
consistent amplitude ii) predicts the resonance peak to remain more
or less at its free space position. Note that the $\Lambda (1405) $
mass shift of the Pauli-blocked amplitude can be reproduced rather
accurately with eq. (\ref{lambda}) and $g_{\Lambda NK}\simeq  2.3
$. We conclude that the attractive feedback effect of a decreased
kaon mass, as included in the self-consistent approach, is
important for the structure of the $\Lambda (1405)$ resonance in
nuclear matter. Altogether the $\Lambda(1405)$ resonance mass is
basically unchanged, however, the resonance acquires an increased
decay width as nuclear matter is compressed. Furthermore we point
out that the real part of the elastic scattering amplitude is
strongly modified: it changes its sign for $\omega > m_N+m_K$ from
repulsion to attraction as the density gets larger.

\begin{figure}[h]
\epsfysize=8cm
\begin{center}
\mbox{\epsfbox{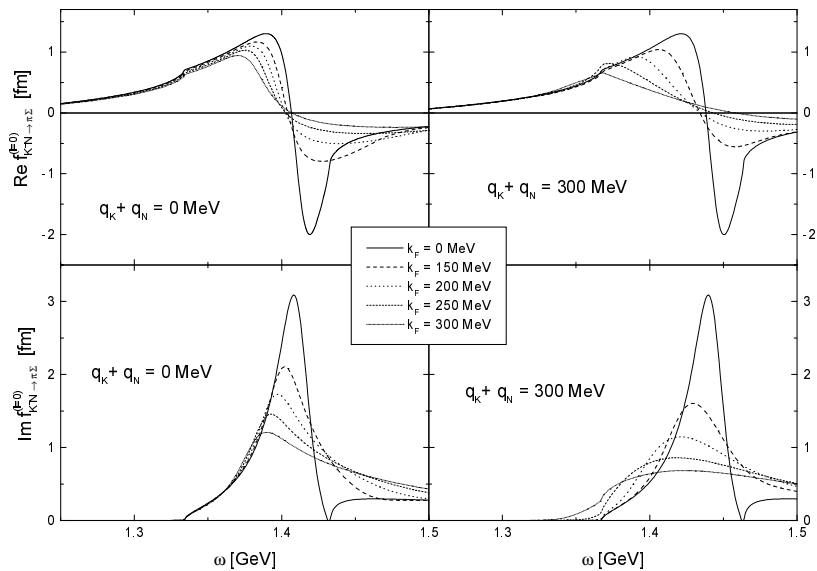}}
\end{center}
\caption{Isospin zero $\pi \Sigma $ production amplitude
in nuclear matter evaluated with self-consistent kaon nucleon
loop.}
\label{fig4}
\end{figure}

We turn to the inelastic channels. Fig. 4 shows the isospin zero
$\pi \Sigma $ production amplitude
\begin{eqnarray}
f^{(I=0)}_{K^-N\rightarrow \pi
\Sigma}(\omega,\vec q )=
\frac{\sqrt{m_N\,m_\Sigma }}{4\,\pi \,\omega } \,
T^{(I=0)}_{K^-N\rightarrow \pi \Sigma } (\omega,\vec q )
\label{}
\end{eqnarray}
evaluated with the self-consistent kaon nucleon loop at $|\vec
q_K+\vec q_N| =0 $ and $|\vec q_K+\vec q_N|=300 $ MeV. The
imaginary part of the amplitude shows a clear peak around 1.4 GeV
for all Fermi momenta representing the $\Lambda(1405) $ resonance
state. This confirms the structure of the resonance as seen in the
elastic channel at $|\vec q_K+\vec q_N|=0 $ MeV. On the other hand,
at the larger momentum $|\vec q_K+\vec q_N|=300 $ MeV the resonance
broadens more quickly and disappears at large Fermi momenta.

We point out that the production amplitude, $f^{(I=0)}_{K^-N
\rightarrow \pi \Sigma }$, is changed considerably close to the kaon nucleon
threshold.
\begin{figure}[h]
\epsfysize=8cm
\begin{center}
\mbox{\epsfbox{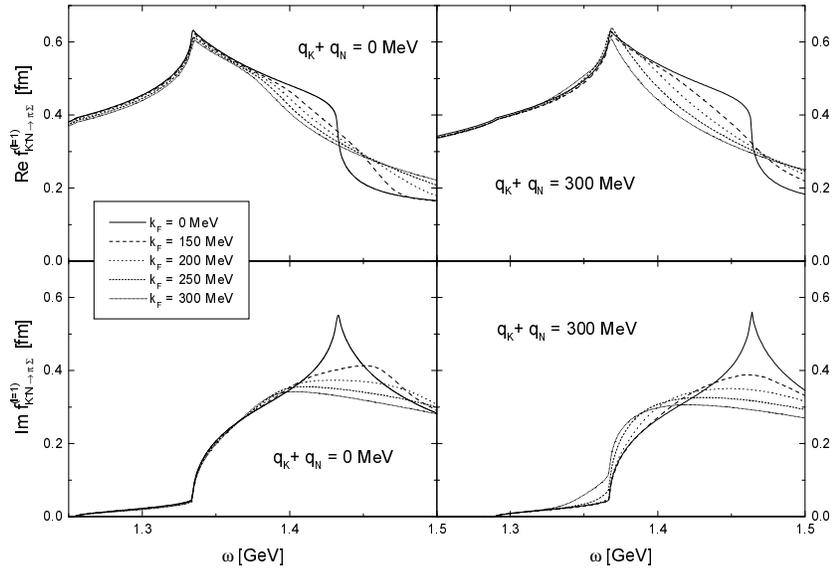}}
\end{center}
\caption{Isospin one $\pi \Sigma $ production amplitude
in nuclear matter evaluated with self-consistent kaon nucleon
loop.}
\label{fig5}
\end{figure}
Finally in Fig. 5 and Fig. 6 we present our result for the isospin
one $\pi \Sigma $ and $\pi \Lambda $ production amplitudes. Again
the kaon nucleon threshold region of the amplitudes is affected
strongly. We conclude that the $\pi \Sigma $ and $\pi
\Lambda $ production branching ratios are modified strongly in nuclear
matter. This should have important consequences for pion and kaon
production in heavy ion collision  at energies where the production
spectra are sensitive to the detailed dynamics of low energy kaon
nucleon scattering \cite{Dover}.

\begin{figure}[h]
\epsfysize=8cm
\begin{center}
\mbox{\epsfbox{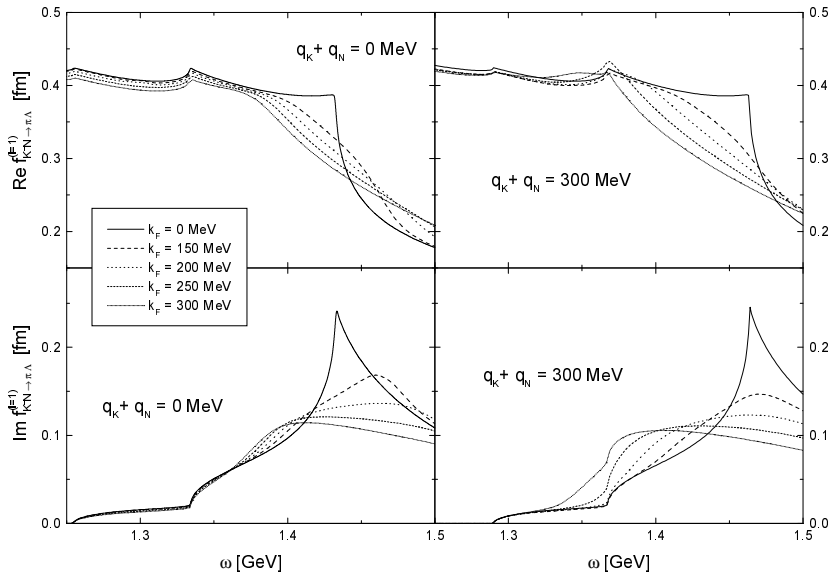}}
\end{center}
\caption{Isospin one $\pi \Lambda $ production amplitude
in nuclear matter evaluated with self-consistent kaon nucleon
loop.}
\label{fig6}
\end{figure}

\section{$K^-$-nuclear optical potential}

Kaonic atom data provide a valuable consistency check on any
microscopic theory of $K^-$ nucleon interaction in nuclear matter.
For a recent review see \cite{report}. It is therefore important to
apply our microscopic approach also to kaonic atoms.

$K^-$ atoms are described in terms of the Klein Gordon equation
\begin{eqnarray}
\Big[\vec \nabla \cdot \vec \nabla -\mu^2+ \Big( \omega-V_c(r) \Big)^2
-2\,\mu\,U_{opt } (\vec r,\vec \nabla )\Big] \phi(\vec r\,) =0
\label{}
\end{eqnarray}
with the Coulomb potential $V_c(r)$ of the finite nucleus including
the proper vacuum polarization and $U_{opt } (\vec r,\vec
\nabla )$ the $K^-$ nuclear optical potential. Here $\mu $ is the
reduced mass of the kaon nucleus system. The binding energy $E$ and
decay width $\Gamma $ of the kaonic atom state follow with $\omega
=\mu+E-i\,\Gamma /2 $.

The optical potential $U_{opt } \left(\vec r,\vec \nabla \right)$
is related to  the $K^-$ self energy in nuclear matter. In the
extreme low density limit the optical potential is determined by
the s-wave $K^-N$ scattering length:
\begin{eqnarray}
2\,\mu\,\,U_{opt } \left(\vec r,\vec \nabla \right) =
-4\,\pi \left( 1+\frac{m_K}{m_N} \right) a_{K^- N}\,\rho (r)
\label{potential1}
\end{eqnarray}
with
\begin{eqnarray}
a_{K^-N} &&=\frac{1}{4}\left(
a_{K^-N}^{(I=0)}+3\,a_{K^-N}^{(I=1)}\right)
\simeq \left( -0.18 +0.67\,i \right) \, \mbox{fm}
\end{eqnarray}
and the nucleus density profile $\rho(r)$.  The optical potential
(\ref{potential1}) does not describe kaonic atoms well. Friedman et
al. show that kaonic atom data can be described with a large
attractive effective scattering length $a_{eff}
\simeq
\left( 0.63 +0.89
\,i\right)  $ fm  which is in direct contradiction with the low
density optical potential (\ref{potential1}).

We compare this large attractive scattering length of Friedman et
al. with  the density dependent effective scattering length
\begin{eqnarray}
\Pi_{K^-} (\omega =m_K , \vec q=0 )
&=&-\frac{8}{3\,\pi} \left(1+\frac{m_K}{m_N} \right)
a_{eff}(k_F)\,k_F^3\;.
\label{aeff}
\end{eqnarray}
predicted by our kaon self energy $\Sigma_{K^-} (\omega,
\vec q\,^2 )$. The real part of the effective scattering length
$a_{eff}(k_F) $, shown in Fig. 7, changes sign as the density is
increased. At large densities we find an attractive scattering
length. This is a welcome effect. However as shown in \cite{wojtek}
this effect falls short in explaining kaonic atom data. A good
description of data requires a large attractive effective
scattering length at $k_F< 150 $ MeV if non-local interaction terms
were small.  We conclude that the explanation of kaonic atoms with
a density dependent effective scattering length but with only small
gradient interaction terms is ruled out by our microscopic
approach.

\begin{figure}[h]
\epsfysize=7cm
\begin{center}
\mbox{\epsfbox{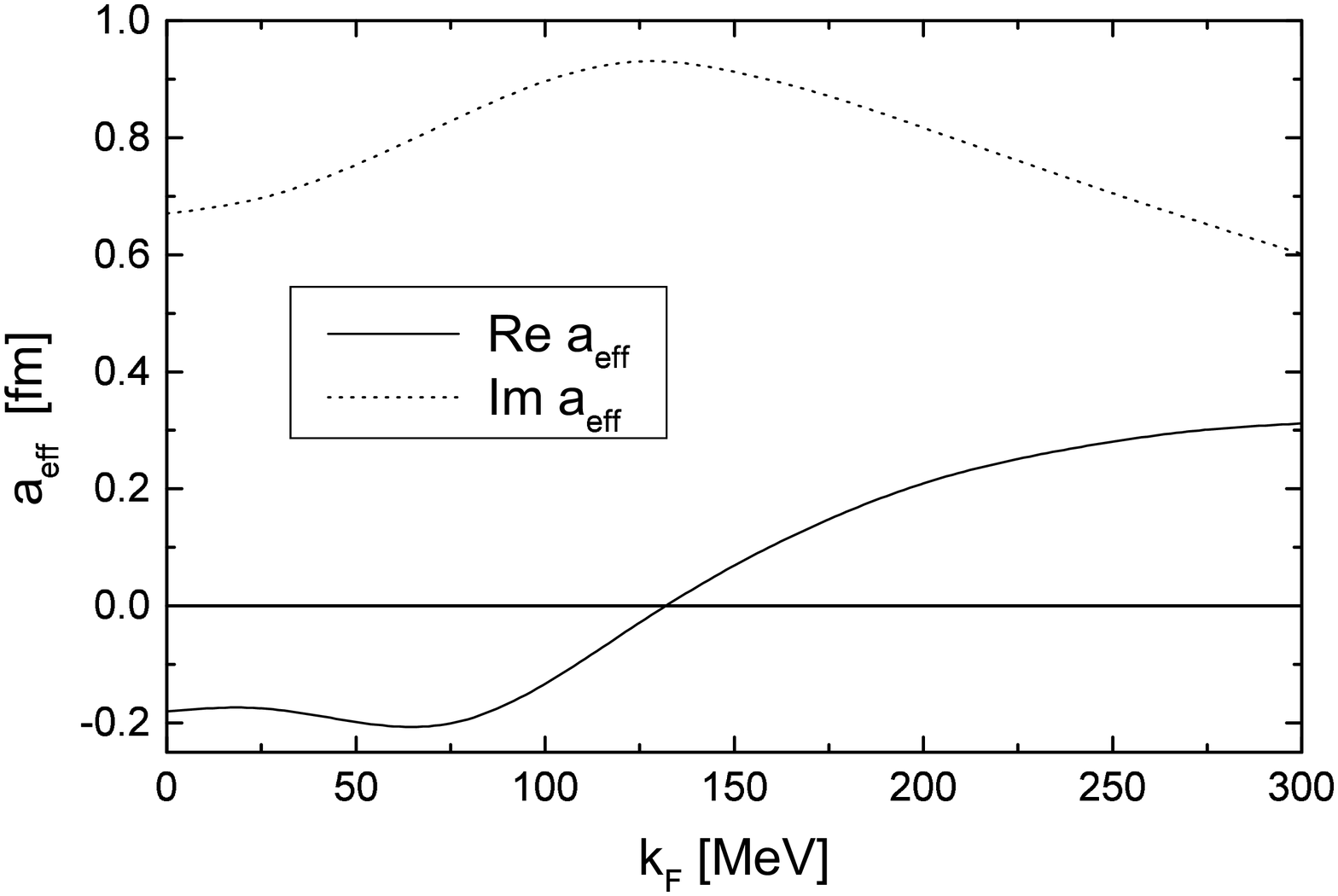}}
\end{center}
\caption{The effective scattering length $a_{eff}(k_F)$ defined
in eq. (\ref{aeff}).}
\label{fig7}
\end{figure}
We proceed and discuss non-local effects. Mizoguchi et al.
\cite{Mizoguchi} use a optical potential of the form
\begin{eqnarray}
2\,\mu\,\,U_{opt } \left(\vec r,\vec \nabla \right) =
-4\,\pi \left( 1+\frac{m_K}{m_N} \right)
\left(a_{K^- N}\,\rho (r)
- b\, \vec \nabla \,\rho(r)\cdot \vec \nabla \right)
\label{potential2}
\end{eqnarray}
with the phenomenological parameter $b \simeq (0.47 +i\,0.30 ) $
fm$^3$ adjusted to reproduce kaonic atom data.

It is instructive to confront the phenomenological parameter $ b$
with our microscopic approach. We extract an effective slope
parameter, $b_{eff}(k_F)$,
\begin{eqnarray}
\Pi_{K^-} (\omega =m_K , \vec q )
&=&-\frac{8}{3\,\pi} \left(1+\frac{m_K}{m_N} \right)
a_{eff}(k_F)\,k_F^3
\nonumber\\
&-&\frac{8}{3\,\pi} \left(1+\frac{m_K}{m_N} \right)
b_{eff}(k_F)\,k^2_F\,\vec q\,^2 +{\cal O}\left( \vec q\,^4 \right)
\label{beff}
\end{eqnarray}
from our self energy with $ |\vec q\,| < k_F $. Naively on may
identify $ k_F\,b=b_{eff}(k_F)$. We point out that the effective
slope parameter, shown in Fig. 8, changes sign at the rather small
Fermi momentum $ k_F\simeq 75 $ MeV. At larger densities our
microscopic approach predicts attractive non-local interaction
strength of considerable size. At a typical Fermi momentum $k_F =
150 $ MeV we find
\begin{eqnarray}
 b \leftrightarrow \frac{1}{k_F}\,b_{eff}(k_F)
\simeq  (0.13+i\,0.13 ) \mbox{fm}^3\;\;\;at \;\;\;k_F =150\;\;\; \mbox{MeV}
\label{}
\end{eqnarray}
a value smaller than the result of Mizoguchi et al.. Together with
the attraction from our effective scattering length and a small
residual attractive p-wave interaction term we expect a good
description of kaonic atom data.

\begin{figure}[h]
\epsfysize=7cm
\begin{center}
\mbox{\epsfbox{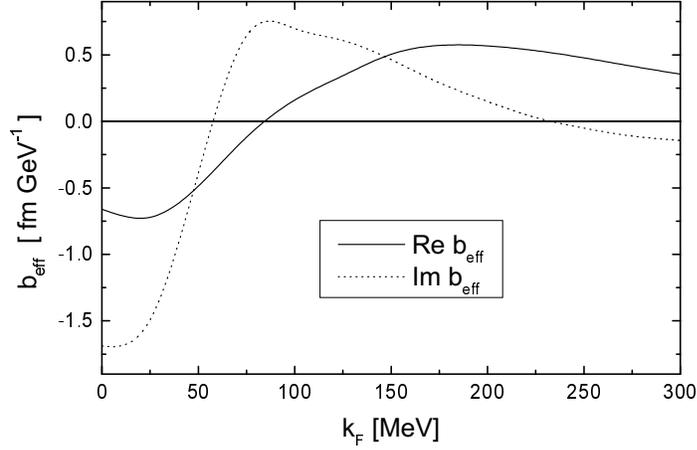}}
\end{center}
\caption{The effective slope parameter $b_{eff}(k_F)$ defined
in eq. (\ref{beff}).}
\label{fig8}
\end{figure}

We point out that the form of the non-local interaction as implied
by our microscopic approach differs from the phenomenological
potential of Mizoguchi et al.. The s-wave $K^-$-nucleon interaction
implies the following form of the optical potential:
\begin{eqnarray}
2\,\mu\,U_{opt } (\vec r,\vec \nabla ) &=&
-4\,\pi \left(1+\frac{m_K}{m_N} \right)
a_{eff}(k_F(r))\,\rho(r)
\nonumber\\
&+&\frac{8}{3\,\pi} \left(1+\frac{m_K}{m_N} \right)
\left(1+\frac{m_K}{2\,m_N}\right)\vec
\nabla \,k^2_F(r)\,b_{eff}(k_F(r))\,\vec \nabla
\nonumber\\
&-&\frac{8}{3\,\pi} \left(1+\frac{m_K}{m_N} \right)
\frac{m_K}{4\,m_N}\Big[ \vec \nabla^2,
k^2_F(r)\,b_{eff}(k_F(r))\Big]_+
\label{koptical}
\end{eqnarray}
with the ordering of the gradients derived in \cite{wojtek}. Our
optical potential (\ref{koptical}) together with a small attractive
p-wave interaction term ( induced by the $Y(1385)$ resonance in the
$K^-n$ channel ) is confronted with kaonic atom data in
\cite{wojtek}.

\section*{Acknowledgments}

The author is grateful for useful discussions with W. Florkowski,
B. Friman, E. Kolomeitsev and D.-P. Min.

\end{document}